\documentclass[12pt,tightenlines,eqsecnum,floats,showpacs,nofootinbib,amsmath,amssymb]{revtex4}

\usepackage{amsmath,amssymb}

\def\be{\begin{equation}}
\def\ee{\end{equation}}
\def\nn{\nonumber}
\def\f{\frac}
\def\tf{\tfrac}

\def\pl{{\rm Pl}}
\def\lp{\ell_\pl}
\def\mC{\mathcal{C}}
\def\mH{\mathcal{H}}

\def\mM{\mathcal{M}}
\def\eff{{\rm eff}}
\def\b{\bar}
\def\d{\dot}
\def\h{\hat}

\def\v{\vec}
\def\wh{\widehat}
\def\dd{{\rm d}}
\def\na{\nabla}
\def\del{\partial}
\def\de{\delta}
\def\ga{\gamma}
\def\la{\lambda}
\def\om{\omega}
\def\ve{\varepsilon}
\def\vp{\varphi}
\def\oe{\mathring{e}}
\def\ow{\mathring{\omega}}
\def\oq{\mathring{q}}

\def\ket{\rangle}

\begin{document}

\title{Holonomy Corrections in the Effective Equations for Scalar Mode Perturbations in Loop Quantum Cosmology}

\author{Edward Wilson-Ewing} \email{wilson-ewing@cpt.univ-mrs.fr}
\affiliation{Centre de Physique Th\'eorique de
Luminy\footnote{Unit\'e mixte de recherche (UMR 6207) du CNRS et
des Universit\'es de Provence (Aix-Marseille I), de la
M\'editerran\'ee (Aix-Marseille II) et du Sud (Toulon-Var);
laboratoire affili\'e \`a la FRUMAM (FR 2291).},
Case 907, F-13288 Marseille, EU}

\begin{abstract}

We study the dynamics of the scalar modes of linear perturbations around a flat,
homogeneous and isotropic background in loop quantum cosmology.  The equations of
motion include quantum geometry effects and are expected to hold at all curvature
scales so long as the wavelengths of the inhomogeneous modes of interest remain
larger than the Planck length.  These equations are obtained by including holonomy
corrections in an effective Hamiltonian and then using the standard variational
principle.  We show that the effective scalar and diffeomorphism constraints are
preserved by the dynamics.  We also make some comments regarding potential inverse
triad corrections.

\end{abstract}


\maketitle

\section{Introduction}
\label{s.intro}

The aim of loop quantum cosmology (LQC) is to use the methods and techniques of loop
quantum gravity in order to study the dynamics of cosmological space-times when the
space-time curvature is of the order of the Planck scale and general relativity breaks
down.  For a recent comprehensive review of LQC, see \cite{as-rev}.  There has been
great progress in LQC over the past few years, especially with regards to the simplest
cosmological models, the homogeneous and isotropic Friedmann-Lema\^itre-Robertson-Walker
(FLRW) space-times.  The dynamics of the quantum theory have been studied extensively,
using both analytical and numerical methods, and it has been found that the big bang
singularity is resolved.  In essence, quantum geometry effects introduce a strong
repulsive force which causes a ``bounce'': there is a quantum gravity bridge between
an earlier classical contracting universe and a later classical expanding universe
\cite{aps, acs}.

Now that the homogeneous sector of LQC is relatively well understood, the next step
is to allow for small perturbations around a homogeneous and isotropic background and
study how the presence of quantum geometry effects modify their dynamics when the
space-time curvature is near the Planck scale.  In this work we will focus on scalar
perturbations; these modes are of particular interest in cosmology as they seed
structure formation.  With the results presented here, it will be possible to study
the dynamics of scalar perturbations through the bounce for modes whose wavelengths
remain larger than the Planck length.

Since these results are of interest to researchers who are not experts in LQC, we have
tried to ensure that this paper is as easy to read for nonspecialists as possible.  If
the reader wishes to skip directly to the main result, the quantum-gravity-corrected
equations for scalar perturbations around a flat homogeneous and isotropic background
(presented in a notation familiar to cosmologists) are given in Sec.\ \ref{ss.eff-fluid}.

In this paper, we will work with effective equations.  Effective equations are obtained
from a Hamiltonian constraint which has been modified in an appropriate way in order
to incorporate quantum gravity effects.  For homogeneous and isotropic models,
effective equations provide an excellent approximation to the full quantum dynamics
of sharply peaked states, even at the bounce point when the quantum gravity effects
are strongest \cite{vt}, and we expect that this will continue to be true if small
perturbations are allowed.

There are two main types of corrections due to quantum
gravity effects in LQC: holonomy corrections and inverse triad corrections.  Holonomy
corrections arise as the connection is expressed in terms of holonomies of a minimal
length and these corrections become important when the space-time curvature nears the
Planck scale.  Inverse triad corrections are introduced in order to obtain well-defined
operators corresponding to inverse powers of the area operator (this is necessary as
0 is in the discrete spectrum of the area operator) and they play an important role
when physical length scales become as small as the Planck length.  We will only consider
holonomy corrections here so that the equations hold for all curvature scales, but
additional inverse triad corrections are necessary if any of the physical length
scales of the space-time become comparable to the Planck length.

Previous work studying cosmological perturbations in LQC at the effective level
has followed a prescription, originally presented in Ref.\ \cite{langlois}, where
the background and perturbation degrees of freedom are separated at the very beginning
in the Hamiltonian constraint and the symplectic structure.  This is certainly a
possible approach to the problem, but it seems to be an unnatural way to try to
incorporate \emph{nonperturbative} quantum gravity corrections.  For this reason, in
this work we will separate the variables into the background and perturbation parts
only after the equations of motion have been determined.  Once it is clear how to
incorporate the holonomy corrections properly, it would of course be possible to go
back and do it again using perturbed variables from the beginning, but the point is
that the problem becomes less mysterious if one works with nonperturbed variables.

Vector and tensor modes were the first to be studied using effective equations in LQC
\cite{bh-v, bh-t, mcbg}.  This was done using perturbed variables as in \cite{langlois}
and both holonomy and inverse triad corrections have been computed.  Scalar modes
have also been studied in \cite{bhks1, bhks2} where inverse triad corrections were
included, and in \cite{als, wl} where holonomy corrections were considered.  Finally,
there have also been some preliminary results about the phenomenology of quantum
gravity effects that could potentially be detected in primordial gravitational wave
signatures \cite{cmns, gcbg} and in the cosmic microwave background \cite{bct, aan}.
As the subtleties of the effective equations are better understood, phenomenological
studies will become more robust and, we hope, will lead to falsifiable predictions.

Scalar perturbations are the most interesting ones from an observational point of view
and therefore it is important to understand how the equations of motion of the scalar
perturbations are modified by quantum geometry effects.  Since in homogeneous models,
where the quantum dynamics are well understood \cite{as-rev, aps, acs}, modifications
due to holonomy corrections are significantly more important than those coming from
inverse triad effects, we might expect this to continue to be true when small perturbations
are allowed.  This is why the study of holonomy corrections to the scalar modes of
cosmological perturbations is important.

While there has been some previous work studying holonomy corrections for scalar
perturbations \cite{als, wl}, there remain many open problems.  In Ref.\ \cite{als},
only large wavelength modes were considered and therefore all spatial derivative
terms in the equations of motion were neglected.  However, it is not clear that the
wavelength of the modes observed today in the cosmic microwave background would
necessarily have been large when the holonomy corrections are important (i.e., when
the curvature nears the Planck scale).  In some scenarios, such as inflation, this
assumption clearly fails.  Because of this, the results presented in \cite{als}
provide only partially corrected equations of motion for scalar perturbations.  More
recently, a further study has appeared where an effective Hamiltonian, including
holonomy corrections, is studied \cite{wl}.   However, the equations obtained from
the effective Hamiltonian constraint are not consistent and one must add an
\emph{ad hoc}, non-unique modification to one of them in order to ensure that the
constraints are preserved by the dynamics.  It is not clear why the chosen modification
should be preferred over any other.  In this work, we will rectify the shortcomings
of the previous studies of holonomy corrections to scalar perturbations by working
with all wavelength modes and by introducing an effective Hamiltonian whose equations
of motion automatically preserve the constraints.

As mentioned above, we will work with nonperturbed variables in the effective
Hamiltonian constraint and symplectic structure.  As we shall see, by doing this it
will become clearer how to incorporate holonomy corrections from an LQC point of view.
However, in order to work with nonperturbed variables in LQC, it is essential to have
a diagonal metric.  If the metric is not diagonal, a key mathematical simplification
is lost: in homogeneous models, it is enough to only consider a small class of
holonomies, called almost periodic functions.  This is a great simplification which is
not available in general, but can be generalized in a straighforward way to include
space-times where the metric can be put in a diagonal form.  Therefore, we will work
in the longitudinal gauge where the metric is diagonal.  This gauge choice is
absolutely necessary if one only wishes to consider almost periodic holonomies: if
one works in a different gauge (or without choosing a gauge), then one must work with
a much larger class of holonomies and the problem becomes much more complicated.  We
will discuss this further in Sec.\ \ref{s.eff}.

The outline of the paper is as follows: in Sec.\ \ref{s.cl}, we present a first
order Hamiltonian formalism for cosmological perturbation theory which gives the
standard equations for the scalar modes in general relativity.  Then in Sec.\
\ref{s.eff} we modify the Hamiltonian in a suitable fashion in order to include
holonomy corrections, this gives the quantum-gravity-corrected equations of motion
for scalar mode perturbations.  We also make some comments about inverse triad
corrections in Sec.\ \ref{s.inv} before ending with a discussion in Sec.\ \ref{s.dis}.

In this paper, we will work in units where $c=1$ but we will keep $G$ and $\hbar$
explicit so that it will be possible to see whether a contribution is due to
gravitational effects, quantum effects or both.  We will define the Planck length
as $\lp = \sqrt{G\hbar}$.  Since we are only considering linear perturbations,
all terms that are second order (or higher) in the perturbations will be dropped
and thus all of the equations are understood to hold up to first order in
perturbations.

\section{First Order Hamiltonian Framework}
\label{s.cl}

In order to study small fluctuations around the flat FLRW cosmological background
on a 3-torus, one allows small departures from homogeneity.  A nice coordinate choice
is the longitudinal gauge in which case the metric is given by%
\footnote{In this work we consider the case of a massless scalar field.  In this
case (and for all other perfect fluids) there is no anisotropic stress, and
therefore the variable $\phi$ encoding the fluctuations around the lapse is
necessarily equal to the variable $\psi$ which describes the deviations
away from the scale factor.  In some cases, this fails to hold in a quantum-corrected
effective setting \cite{bhks2}, but as we shall see the equations we obtain at the
end of the following analysis are consistent and it does not appear to be necessary
to allow for departures from $\phi = \psi$ here.}
%
\be \label{metric} \dd s^2 = -(1 + 2 \psi) \dd t^2 + a^2 (1 - 2 \psi) \dd \v x^2, \ee
where $a(t)$ is the scale factor and depends only on time while $\psi(\v x, t)$
encodes the fluctuations away from the mean scale factor and varies both with time
and position.  We have chosen the line element so that the volume of the 3-torus,
with respect to the background metric given by $\dd \mathring{s}^2 = \dd \v x^2$,
is 1.  Of course, one is free to make a different choice and one can check that
this choice does not affect the results of this work.

Let us justify the choice of the longitudinal gauge.
This gauge choice is particularly useful as the resulting metric is diagonal and this
considerably simplifies the situation as it will be possible to make a direct analogy
with the homogeneous case without first perturbing the Hamiltonian constraint.  In
effect, a diagonal metric allows us to restrict our attention to holonomies that are
almost periodic in the connection.  It is precisely this simplification that has
allowed the vast amount of progress in LQC over the past few years.  This is also why
it is difficult to include perturbations.  Typically this means working with
off-diagonal terms in the metric, which must either be treated in a perturbative manner,
or one must consider a more general class of holonomies than almost periodic functions
of the connection.  However, as we shall see in Sec.\ \ref{s.eff} the longitudinal gauge
allows us to avoid these two problems in the case of scalar perturbations.

\subsection{Elementary Variables}
\label{ss.var}

In order to use a first order formalism, it is necessary to work with triads and co-triads
instead of a metric.  In the longitudinal gauge, the co-triads are given by
\be \om_a^i = a(1-\psi)\ow_a^i, \ee
and the spatial metric is then given by $q_{ab} = \om_a^i \om_{bi}$.  The space-time indices
$a, b, c, \ldots$ are raised and lowered by $q_{ab}$ while the internal indices $i, j, k, \ldots$
are raised and lowered by $\de_{ij} = {\rm diag} (1, 1, 1)$.  Similarly, their inverse the triads
are given by
\be e^a_i = \f{1}{a}(1+\psi)\oe^a_i. \ee
Note that since $\psi \ll 1$, we can drop all terms of the order of $\psi^2$ or higher.
The fiducial triads and co-triads used in the equations above are defined as
\be \oe^a_i = \left(\f{\del}{\del x^i}\right)^a, \qquad \ow_a^i = (\dd x^i)_a. \ee

There exists a derivative operator $D$ compatible with the triads and co-triads,
\be D_a e^b_i = \del_a e^b_i + \Gamma^b_{ac} e^c_i + \epsilon_{ij}{}^k \Gamma_a^j
e^b_k = 0, \ee
where $\Gamma^b_{ac}$ is the usual Christoffel connection and $\Gamma_a^i$ is the
spin-connection,
\be \Gamma_a^i = -\epsilon^{ijk} e^b_j \left(\del_{[a} \om_{b]k} + \f{1}{2} e^c_k \om_a^l
\del_{[c} \om_{b]l} \right). \ee
The Christoffel symbols will not be used in this work, but it is necessary to calculate
the spin-connection,
\be \Gamma_z^2 = -\Gamma_y^3 = \ve \del_x \psi, \quad
\Gamma_x^3 = -\Gamma_z^1 = \ve \del_y \psi, \quad
\Gamma_y^1 = -\Gamma_x^2 = \ve \del_z \psi. \ee
Here $\ve = \epsilon^{123}$ which can be $\pm1$ and therefore $\ve^2 = 1$.

Now, in order to study the perturbations in the (classical) loop gravity framework, it is
necessary to use the Ashtekar connection and densitized triads as our basic variables.
Since the metric is diagonal, we can parametrize the densitized triads by
\be E^a_i = p \sqrt{\oq} \oe^a_i, \qquad {\rm where} \qquad p = a^2 (1 - 2 \psi). \ee
The parameter $p$ is a function of position and time, but we will drop the arguments in
order to simplify the notation, except where they are essential.

The Ashtekar connection is given by $A_a^i = \Gamma_a^i + \ga K_a^i$, where $\ga$ is
the Barbero-Immirzi parameter and $K_a^i = K_{ab} e^{bi}$ is related to the extrinsic
curvature.  It is easy to show that, like $E^a_i$, $K_a^i$ is diagonal (with respect
to the fiducial triads) and all of its entries are equal:
\be \label{Kab} K_a^i = (\d a - 2 \d a \psi - a \d\psi) \ow_a^i. \ee
A useful property of the Ashtekar connection in this context is that its diagonal terms
solely come from $K_a^i$, while its off-diagonal terms solely come from $\Gamma_a^i$ (of
course, this is not the case in general).  Since the densitized triads are diagonal, only
the diagonal part of the Ashtekar connection will appear in the induced symplectic
structure and therefore it is convenient to parametrize the Ashtekar connection by
\begin{align} A_x &= c \, \tau_1 - \ve (\del_z \psi) \, \tau_2 + \ve (\del_y \psi) \, \tau_3,
\nn \\ A_y &= \ve (\del_z \psi) \, \tau_1 + c \, \tau_2 - \ve (\del_x \psi) \, \tau_3,
\\ A_z &= -\ve (\del_y \psi) \, \tau_1 + \ve (\del_x \psi) \, \tau_2 + c \, \tau_3, \nn
\end{align}
where $A_a = A_a^i \tau_i$ and the $\tau_i$ are a basis of the Lie algebra of SU(2) such
that $\tau_i \tau_j = \tf{1}{2} \epsilon_{ij}{}^k \tau_k - \tf{1}{4} \de_{ij} \mathbb{I}$.
Following this definition, we find that the induced symplectic structure on our phase
space gives the following nonzero Poisson bracket:
\be \label{poisson} \{ c(\v x), p(\v y) \} = \f{8 \pi \ga G}{3} \de^3(\v x - \v y). \ee

\subsection{Massless Scalar Field}
\label{ss.phi}

Since we are primarily interested in the effects due to quantum gravity in the early
universe, we will take the simplest matter field possible, a massless scalar field.
The action for a massless scalar field is
\be S = -\f{1}{2} \int_\mM \sqrt{|g|} \, g^{\mu \nu} \left( \partial_\mu \vp \right)
\left( \partial_\nu \vp \right), \ee
and it is easy to show that the conjugate momentum to $\vp$ is given by
\be \label{defpi} \pi_\vp = N \sqrt{|q|} \d \vp, \ee
where the dot represents derivation with respect to time and $N$ is the lapse function.
The Poisson bracket is given by
\be \{ \vp(\v x), \pi_\vp(\v y) \} = \de^3(\v x - \v y). \ee

In this work, we will be using the Hamiltonian formulation of general relativity and
therefore the stress energy tensor will not appear.  Since a lot of the literature
in the field of cosmological perturbation theory starts from the Einstein equations
(and thus uses the stress energy tensor), the following dictionary can be useful in
order to compare the results given here with those available in the literature.

For a massless scalar field, the stress energy tensor is given by
\be \label{tmunu} T^\mu{}_\nu = g^{\mu\la} \left( \partial_\la \vp \right)
\left( \partial_\nu \vp \right) - \f{1}{2} \de^\mu_\nu \left( \v\na \vp \right)^2. \ee
For small perturbations around a flat FLRW background, we have $\vp = \b \vp + \de \vp$
where the bar denotes the homogeneous background and $\de \vp$ is the inhomogeneous
perturbation.  One can immediately see that the second term in $T^\mu{}_\nu$ is
negligible as it is second order in $\de \vp$.

The degrees of freedom of a perfect fluid are given by the energy density $\rho$, the
pressure $P$ and the velocity four-vector $u^\mu$.  The energy density and pressure can
be split into background quantities and perturbations, $\rho = \b \rho + \de \rho,
P = \b P + \de P$, while the only scalar mode of the perturbation of the four-velocity
at first order is given by $\de u$, which affects the spatial part of the four-velocity
by $\de u_a = \del_a (\de u)$.  In terms of these variables, the stress energy tensor
is given by (see, e.g., \cite{weinberg, mukhanov})
\begin{align} T^0{}_0 &= -\b\rho - \de \rho, \\
T^0{}_a &= (\b\rho + \b P) \partial_a (\de u), \\
T^a{}_b &= \de^a_b (\b P + \de P), \end{align}
and then by Eq.\ \eqref{tmunu} the relations between the ``standard'' variables and the
ones used in the Hamiltonian framework are the following:
\be \label{defbrho} \b\rho = \b P = \tf{1}{2} \d{\b\vp}^2, \ee
\be \label{defdrho} \de\rho = \de P = \d{\b\vp} \, \d{(\de \vp)} - \d{\b\vp}^2 \, \psi, \ee
\be \label{defdu} (\b\rho + \b P) \de u = - \d{\b\vp} \, \de \vp. \ee
Another useful relation can be obtained by adding Eqs.\ \eqref{defbrho} and \eqref{defdrho},
this gives
\be \label{defrho} \rho = P = (1 - 2 \psi) \f{\d \vp^2}{2}. \ee

\subsection{The Hamiltonian Constraint and Dynamics}
\label{ss.dyn}

Before writing the Hamiltonian constraint, it is necessary to express $\psi$ in terms of
the conjugate variables $(c, p)$.  In order to do this, we will assume that $\int_\mM
\psi = 0$ (i.e., the zero mode of the Fourier decomposition of $\psi$ is zero) and then
it follows that
\be \label{psipbar} \psi = \f{\b p - p}{2 \b p}, \quad {\rm where} \quad \b p = \int_\mM p. \ee
This requires the introduction of the nonlocal $\b p$, but it is necessary in order to
express the perturbation $\psi$ in terms of the elementary variable $p$.  As we shall
see, the resulting nonlocal Hamiltonian system provides the usual local equations of
motion for linear cosmological perturbations.

The Hamiltonian constraint is given by
\be \mC_H = \int_\mM \big[ N \mH + N^a \mH_a + N^i \mathcal{G}_i \big], \ee
where, for a massless scalar field, the scalar constraint is
\be \mH = \f{-E^a_i E^b_j}{16 \pi G \gamma^2 \sqrt{|q|}} \epsilon^{ij}{}_k \bigg( F_{ab}{}^k
- (1 + \ga^2) \Omega_{ab}{}^k \bigg) + \f{1}{2\sqrt{|q|}} \bigg( \pi_\vp^2 + E^a_i E^{bi} (\del_a
\vp) (\del_b \vp) \bigg) \approx 0, \ee
the diffeomorphism constraint is
\be \mH_a = \f{E^b_i}{4 \pi G \ga} F_{ab}{}^i + \pi_\vp \partial_a \vp \approx 0, \ee
and the Gauss constraint is
\be \mathcal{G}_i = \mathcal{D}_a E^a_i = \del_a E^a_i + \epsilon_{ij}{}^k A_a^j E^a_k \approx 0. \ee
The $F_{ab}{}^k$ appearing above is field strength of the Ashtekar connection,
\be F_{ab}{}^k = 2 \del_{[a} A_{b]}^k + \epsilon_{ij}{}^k A_a^i A_b^j. \ee
Similarly, $\Omega_{ab}{}^k$ is the field strength of the spin-connection.

The $\approx 0$ indicates that $\mH, \mH_a$ and $\mathcal{G}_i$ are constraints and must
vanish for physical solutions.  A solution to these three constraints at an initial time can
then be evolved to later times by taking Poisson brackets with the Hamiltonian constraint
$\mC_H$.  An important point is that any solution that initially satisfies the scalar,
diffeomorphism and Gauss constraints will continue to do so under the evolution determined
by $\mC_H$.

For linear perturbations around the flat FLRW space-time in the longitudinal gauge, we find
that the Gauss constraint is automatically satisfied, while the scalar constraint becomes
\be \label{ham-cl} \mH = \sqrt{\oq} \Bigg[\f{-\sqrt{|p|}}{8 \pi G} \bigg( \f{3 c^2}{\ga^2}
+ 2 \na^2 \left( \tf{\b p - p}{2 \b p} \right) - \left( \v\na \tf{\b p - p}{2 \b p} \right)^2
\bigg) + \f{\pi_\vp^2}{2 p^{3/2} \oq} + \f{\sqrt{p}}{2} \left( \v\na \vp \right)^2 \Bigg]
\approx 0, \ee
and the diffeomorphism constraint is given by
\be \label{diff-cl} \mH_a = \f{\sqrt{\oq} p}{4 \pi G \ga} \bigg[ \del_a c
+ c \, \del_a \left( \tf{\b p - p}{2 \b p} \right) \bigg]
+ \pi_\vp \del_a \vp \approx 0. \ee
Finally, since $N = 1 + \psi = 1 + \tf{\b p - p}{2 \b p}$ and $N^a = 0$ in the longitudinal
gauge, the Hamiltonian constraint is given by
\be \label{ch-cl} \mC_H = \int_\mM \bigg( 1 + \tf{\b p - p}{2 \b p} \bigg) \mH.
\ee
Although one might be tempted to drop terms in $\mH$ like $(\v\na \vp)^2$ which are second order
in the perturbations, they contribute first order terms in the evolution equations (in this case
for $\d \pi_\vp$) after the derivative is integrated by parts.

This Hamiltonian formalism is of course well-defined for all $c(\v x), p(\v x), \vp(\v x)$ and
$\pi_\vp(\v x)$, but the only case we are interested in is a homogeneous background with small
perturbations in which case the resulting system is equivalent to linear cosmological perturbations
in general relativity.  Therefore, we will only consider equations for linear perturbations around
a homogeneous background in what follows.

It is now possible to obtain the standard results for linear perturbations in cosmology from
the two constraints and the Hamiltonian.  Using the relations (but only after the equations of
motion have been obtained by the variational principle)
\be p = a^2 (1 - 2\psi),\: \b p = a^2,\: c = \b c + \de c,\: \vp = \b \vp + \de \vp,\:
\pi_\vp = \b \pi_\vp + \de \pi_\vp, \ee
where as usual the ``barred'' quantities correspond to the unperturbed background quantities,
we find that the scalar constraint $\mH = 0$ implies that
\be -\f{a c^2}{\ga^2}(1 - \psi) - \f{2a}{3} \na^2 \psi + \f{4 \pi G}{3} \f{\pi_\vp^2}{p^{3/2}\oq}
= 0, \ee
where we have only kept terms linear in the perturbations.  The diffeomorphism constraint
$\mH_a = 0$ gives
\be \sqrt{\oq}\, \f{a^2}{\ga} \Big[ \del_a(\de c) + \b c \, \del_a \psi \Big] + 4 \pi G \pi_\vp
\del_a (\de \vp) = 0. \ee

The dynamics are obtained from the Hamiltonian constraint.  Starting with the equations of
motion for the matter degrees of freedom since they are simpler,
\begin{align} \label{phidot-cl} \d \vp &= \{ \vp, \mC_H \} = \f{\de \mC_H}{\de \pi_\vp}
= (1 + \psi) \f{\pi_\vp}{p^{3/2} \sqrt{\oq}}, \\ \label{pidot-cl} \d \pi_\vp
&= \{ \pi_\vp, \mC_H \} = -\f{\de \mC_H}{\de \vp} = \sqrt{\oq}\, a \na^2 (\de \vp). \end{align}
Note that the first relation is equivalent to Eq.\ \eqref{defpi}, as one should expect and
it also implies, together with Eq.\ \eqref{defrho}, that
\be \label{dyn-rho} \rho = \f{\pi_\vp^2}{2 |p|^3 \oq}, \ee
which will be a useful relation later.

The equation of motion for $\d p$ also has a simple form,
\be \label{pdot-cl} \d p = \{ p, \mC_H \} = -\f{8 \pi \ga G}{3} \f{\de \mC_H}{\de c}
= \f{2}{\ga} (1 + \psi) \sqrt{|p|} c, \ee
which, solving for $c$, gives
\be \label{c-cl} c = \ga \left( \d a - 2 \d a \psi - a \d \psi \right), \ee
which is exactly what is given in Eq.\ \eqref{Kab} (recall that the diagonal portion
of $A_a^i$ is given by $\ga K_a^i$ as the diagonal part of $\Gamma_a^i$ is zero).

Finally the equation of motion for $\d c$ is the most complicated as one must vary
$\mC_H$ with respect to $p$.  There are two contributions that must be integrated by
parts (since the topology is $T^3$ there are no boundary terms) and there are also terms
which depend on $\b p$ which could potentially contribute.  Since $\de \b p / \de p (\v x)
= 1$ ---note that there is no delta function--- all of the terms that are obtained by
varying $\b p$ with respect to $p (\v x)$ are inside an integral over $\mM$ and would
contribute a nonlocal term to the dynamics; this would disagree with the well known
general relativity results.  However, all of the nonlocal terms vanish and one is left
with a local equation of motion.  In order to see this, we start with
\be \d c(\v{x}) = \{ c(\v{x}), \mC_H \} 
= \f{8 \pi \ga G}{3} \int_\mM \dd^3 \v{y} \left[ \f{\de \psi(\v{y})}{\de p(\v{x})} \mH(\v{y})
+ [1 + \psi(\v{y})] \f{\delta \mH(\v{y})} {\de p(\v{x})} \right], \ee
where the first term can be dropped as it is multiplied by the constraint $\mH$ which
vanishes on-shell.  A straightforward calculation then gives
\begin{align} \d c(\v{x}) =& - \f{[1 + \psi(\v{x})] c(\v{x})^2}{2 \ga \sqrt{p(\v{x})}}
- \f{\ga}{3a} \na^2 \psi(\v{x}) - \f{2 \pi \ga G [1 + \psi(\v{x})] \pi_\vp(\v{x})^2}
{p(\v{x})^{5/2}} \nn \\ & \label{cdot-st1}
- \f{2 \ga a}{3} \int_\mM \! \dd^3 \v{y} \: \na^2 \f{\de \psi(\v{y})}{\de p(\v{x})}
+ \f{2 \ga a}{3} \int_\mM \! \dd^3 \v{y} \: [\v{\na} \psi(\v{y})] \:
\v{\na} \left( \f{\de \psi(\v{y})} {\de p(\v{x})} \right). \end{align}
The fourth term can dropped as it is a total divergence, while the fifth term requires
a little more work.  From Eq.\ \eqref{psipbar}, it follows that
\be \f{\de \psi(\v{y})} {\de p(\v{x})} = \f{p(\v{y})}{2 \b p^2}
- \f{\de^3(\v{x}-\v{y})}{2 \b p}, \ee
where the delta function only appears in one of the two terms.  Using this relation,
integrating by parts and dropping terms that are second order in $\psi$, the
fifth term in Eq.\ \eqref{cdot-st1} becomes
\be \f{2 \ga a}{3} \int_\mM \! \dd^3 \v{y} \: [\v{\na} \psi(\v{y})] \:
\v{\na} \left( \f{\de \psi(\v{y})} {\de p(\v{x})} \right) = \f{\ga}{3a} \na^2 \psi(\v{x})
- \f{\ga}{3a} \int_\mM \dd^3 \v{y} \: \na^2 \psi(\v{y}).  \ee
The second term vanishes as it is a total divergence and one is left with the local
relation
\be \label{cdot-cl} \d c = - \f{1}{2 \ga a} (1 + 2\psi) c^2
- 2 \pi G \ga a \f{\pi_\vp^2}{|p|^3 \oq}. \ee

In order to compare our results to the standard results in the literature, one can remove
all instances of $\pi_\vp$ and $c$ via Eqs.\ \eqref{phidot-cl} and \eqref{c-cl} and then use
the definitions of $\de \rho$ and $\de u$ given in Eqs.\ \eqref{defdrho} and \eqref{defdu}.
After doing this, the scalar constraint gives
\be \label{ham-std} \f{\d a^2}{a^2} = \f{8 \pi G}{3} \b \rho + \f{8 \pi G}{3} \de \rho
+ 2 \f{\d a^2}{a^2}\psi + 2 \f{\d a}{a} \d \psi - \f{2}{3a^2} \na^2 \psi, \ee
and the diffeomorphism constraint $\mH_a = 0$ becomes
\be \label{diff-std} \del_a \left( \d a \psi + a \d \psi \right)
+ 8 \pi G a \b \rho \del_a (\de u) = 0. \ee
Equation \eqref{pidot-cl} multiplied by $\d \vp$ gives
\be \d{\b \rho} + 6 \f{\d a}{a}\b\rho + \d{\de\rho} + 6 \f{\d a}{a} \de\rho - 6 \b\rho \d\psi
+ \f{2}{a^2}\b\rho \na^2 (\de u) = 0, \ee
while an additional relation between the perturbations in the massless scalar field can be
obtained by taking the time derivative of Eq.\ \eqref{defdu} which gives
\be \label{dudot-cl} \del_t \left( 2 \b\rho \de u \right) = -\de\rho - 2 \b\rho \psi
- 6 \f{\d a}{a} \b\rho \de u, \ee
where the relation $\ddot{\b\vp} = -3 \d a \d{\b \vp} / a$ ---obtained from Eqs.\ \eqref{phidot-cl}
and \eqref{pidot-cl} and the relation $\d{ \b p} = 2 a \d a$--- was used.  Finally, using
Eq.\ \eqref{c-cl}, Eq.\ \eqref{cdot-cl} gives
\be \f{\ddot a}{a}  + \f{\d a^2}{2 a^2} = - 4 \pi G \b \rho - 4 \pi G \de \rho + 2 \f{\ddot a}{a} \psi
+ \f{\d a^2}{a^2} \psi + 4 \f{\d a}{a} \d \psi + \ddot \psi. \ee
These five equations provide the standard results for linear perturbations around the flat FLRW
background (compare with, e.g., \cite{weinberg, mukhanov}) when the matter content is a
massless scalar field; that is, when there is no anisotropic stress and the equation of state
is $P = \rho$.  This shows that the Hamiltonian \eqref{ch-cl} with the constraints \eqref{ham-cl}
and \eqref{diff-cl} gives the correct dynamics for scalar perturbations in cosmology.

\section{Holonomy Corrections}
\label{s.eff}

In this section, we will appropriately modify the scalar constraint in order to incorporate quantum
gravity effects that come from using holonomies of the connection ---rather than the connection
itself--- in the quantum theory.  On the other hand, the diffeomorphism and Gauss constraints are
not modified: this is because they encode symmetries which we expect to remain present at all
scales, including the Planck scale.

The idea is to model the most important quantum gravity effects coming from a quantum theory in
a simpler (often called ``quantum-corrected'') setting which is treated in a classical manner.
One example of a quantum theory that could generate these quantum-corrected equations is a form
of lattice LQC where each cell is described by a homogeneous and isotropic geometry but where
the geometries vary from one cell to another so that inhomogeneities are present at scales
larger than the cell size \cite{boj-inh, we-inh}.  In this setting, the Hamiltonian would be
separated into an ultralocal term (the ``homogenous'' part) and all of the terms containing
derivatives would be treated as interactions between different cells.  This would of course
constitute an approximation but would be valid so long as the volumes of the individual cells
remain larger than the Planck volume and that there are enough cells in the lattice in order
to properly describe the inhomogeneities up to a prescribed maximum wave number.

In the longitudinal gauge, the homogeneous metric in each cell is isotropic and thus the
symplectic structure in each cell would be that of an FLRW space-time.  This indicates that one might
be able to quantize each cell following the methods used in homogeneous models \cite{as-rev, aps, acs}
and then turn on interactions in the Hamiltonian.  It is worth pointing out that the diagonal
form of the metric in the longitudinal gauge will greatly simplify the quantization process.
This is mainly because, as one can easily check, when the metric is nondiagonal the field
strength obtained by taking a holonomy around a square loop of nonzero area is generically not
almost-periodic in the connection (which is proportional to the extrinsic curvature in the
spatially flat homogeneous case where the spin-connection vanishes).  However, in this lattice
LQC quantization, due to the diagonal form of the metric it is possible to restrict our attention
to almost periodic functions of the connection, just as in homogeneous space-times.

Once the quantum theory has been constructed,
the effective theory can be obtained by taking a sharply peaked wave function and studying the
evolution of the expectation values of certain observables under the action of the Hamiltonian
operator.  Some observables such as $p$ can be obtained directly from expectation values of the
fundamental operators in the quantum theory while others ---such as $c$ which one could associate
to the expectation value of an operator corresponding to a short holonomy (minus the identity)
divided by the length of the holonomy--- would require a little more care.  In any case, we will
not construct the quantum theory here but leave it for future research.  Instead, we will build
the effective Hamiltonian by applying the heuristic methods following from (and well supported
by) homogeneous LQC.

In the first part of this section, we will implement holonomy corrections in the system studied in
the previous section, that of a massless scalar field.  Then, in the second part we will present
a conjectured generalization of these results for all perfect fluids.

\subsection{Effective Equations for a Massless Scalar Field}
\label{s.eff-phi}

One of the key parts in the process of the LQC quantization of the Hamiltonian constraint operator
is that the field strength $F_{ab}{}^k$ of the Ashtekar connection must be expressed in terms of a
holonomy of the connection $A_a^i$ around the smallest possible loop.  It is posited that the
smallest loop has an area given by $\Delta \lp^2 = 4 \sqrt{3} \pi \ga \lp^2$, the smallest eigenvalue
of the area operator in loop quantum gravity.

For homogeneous and isotropic cosmologies, it has been found that the full LQC dynamics of states which,
at some initial time are sharply peaked around a classical solution, remain sharply peaked as they are
evolved by the action of the Hamiltonian constraint operator \cite{aps}.  In addition, the dynamics of
the wave packet are extremely well approximated by a modified classical Hamiltonian constraint where one
includes holonomy corrections by determining the field strength by taking a holonomy around the minimal
loop as prescribed above \cite{vt}.  In this sense, it seems that the most important quantum gravity
modifications to the Einstein equations in this very simple setting come from the holonomy corrections.
These corrections become important when the space-time curvature approaches the Planck scale, but are
completely negligible when the matter energy density is less than $0.01 \rho_{\rm Pl}$.

Unfortunately, \emph{a priori} it is not entirely clear how to incorporate holonomy corrections for
inhomogeneous space-times.  In previous works \cite{bh-v, bh-t, mcbg, wl}, generic holonomy correction
terms were added to the effective equations by hand and then the form of the correction terms was
restricted by the condition that the constraint algebra must continue to close.  However, so far it
has proven to be impossible to complete this programme for holonomy corrections to scalar perturbations.

Here, due to the choice of both the gauge and of the variables, the situation is luckily quite
simple.  In the classical scalar constraint, we find that the term with the difference between
the field strengths of the Ashtekar connection and the spin-connection can be simplified,
\be E^a_i E^b_j \epsilon^{ij}{}_k [F_{ab}{}^k - (1 + \ga^2) \Omega_{ab}{}^k] =
E^a_i E^b_j \epsilon^{ij}{}_k [F^{(\rm iso)}_{ab}{}^k - \ga^2 \Omega_{ab}{}^k], \ee
where we have defined 
\be F^{(\rm iso)}_{ab}{}^k = c^2 \epsilon^k{}_{ij} \ow_a^i \ow_b^j, \ee
where of course $c$ depends on position.  Although this ``field'' strength is also inhomogeneous,
it nonetheless has a very similar form to the field strength in homogeneous settings.  Note that
this provides a natural separation of the $F_{ab}{}^k$ and $\Omega_{ab}{}^k$ terms into an
ultralocal term, $F^{(\rm iso)}_{ab}{}^k$, and an interaction term, $\Omega_{ab}{}^k$.  Therefore,
following the arguments introduced at the beginning of this section, we can introduce holonomy
corrections to the ultralocal term $F^{(\rm iso)}_{ab}{}^k$ as in homogeneous models while
$\Omega_{ab}{}^k$ is treated as an interaction term which only depends on $p$ and therefore
will not be modified.  Thus, in order to incorporate holonomy corrections in the scalar
constraint, we shall mimic the procedure in the homogeneous case and replace%
\footnote{The Belinskii, Khalatnikov, Lifshitz (BKL) conjecture ---which suggests that spatial
derivatives can be ignored when the spatial curvature becomes very large \cite{bkl1, bkl2}---
also supports this way of introducing the holonomy corrections.  Since holonomy corrections are
only important when the curvature approaches the Planck regime, if the BKL conjecture holds near
the Planck scale then spatial derivatives and thus $\Omega_{ab}{}^k$ can be ignored when one
incorporates holonomy corrections.  Then only $F^{(\rm iso)}_{ab}{}^k$ remains and the holonomy
corrections are introduced as in Eq.\ \eqref{subF}.  Note that although the BKL conjecture says
nothing about isotropy, $F^{(\rm iso)}_{ab}{}^k$ is locally isotropic due to the locally isotropic
form of the metric in the longitudinal gauge.}
%
\be \label{subF} F^{(\rm iso)}_{ab}{}^k = c^2 \epsilon^k{}_{ij} \ow_a^i \ow_b^j \to
\f{\sin^2 \b\mu c} {\b\mu^2} \epsilon^k{}_{ij} \ow_a^i \ow_b^j, \ee
where
\be \b\mu = \sqrt\f{\Delta \lp^2}{|p|}, \ee
just as in the homogeneous case \cite{as-rev, aps}, except that $p = a^2 (1 - 2 \psi)$ is no
longer homogeneous.

Thus, the scalar constraint becomes
\begin{align} \label{scal-hol} \mH^\eff = \sqrt{\oq} & \Bigg[\f{-1}{8 \pi G} \bigg(
\f{3 |p|^{3/2}}{\ga^2 \Delta \lp^2} \sin^2 \b\mu c
+ 2 \sqrt{|p|} \na^2 \left( \tf{\b p - p}{2 \b p} \right)
- \sqrt{|p|} \left( \v\na \tf{\b p - p}{2 \b p} \right)^2 \bigg) \nn \\ 
& + \f{\pi_\vp^2}{2 |p|^{3/2} \oq}
+ \f{\sqrt{|p|}}{2} \cos 2 \b\mu c \left( \v\na \vp \right)^2 \Bigg]
\approx 0, \end{align}
where the $\cos 2 \b\mu c$ in the last term is added in order to ensure that the effective scalar
constraint is preserved by the dynamics, i.e., that $\d \mH^\eff \approx 0$.

On the other hand, the diffeomorphism constraint is not modified:
\be \mH_a^\eff = \f{\sqrt{\oq} p}{4 \pi G \ga} \Big[ \del_a c
+ c \del_a \left(\tf{\b p - p}{2 \b p}\right) \Big] + \pi_\vp \del_a \vp \approx 0. \ee
The Gauss constraint is not modified either and continues to hold due to the choice of the
longitudinal gauge and therefore we do not need to consider it any further.  As an aside, we
point out that it is not immediately clear how to incorporate an unchanged diffeomorphism
constraint in lattice LQC, we leave this problem for future work.

As before, the Hamiltonian constraint provides the dynamics and is simply given by
\be \mC_H^\eff = \int_\mM \left( 1 + \tf{\b p - p}{2 \b p} \right) \mH^\eff. \ee
The scalar and diffeomorphism constraints must be satisfied by the initial data and
thus
\be \label{sc-eff} - \f{|p|^{3/2}}{\ga^2 \Delta \lp^2} \sin^2 \b\mu c
- \f{2 a}{3} \na^2 \psi + \f{4 \pi G}{3 |p|^{3/2} \oq} \pi_\vp^2 = 0, \ee
\be \label{dc-eff} \sqrt{\oq} \f{p}{4 \pi G \ga} \left[ \del_a (\de c) + c \del_a \psi \right]
+ \pi_\vp \del_a (\de \vp) = 0. \ee
Finally, the equations of motion generated by $\mC_H^\eff$ are the following:
\begin{align}
\label{phidot-eff} \d \vp & = (1 + \psi) \f{\pi_\vp}{p^{3/2} \sqrt{\oq}}, \\
\label{pidot-eff} \d \pi_\vp & =  \sqrt{\oq}\, a \cos 2 \b\mu c \, \na^2 \vp, \\
\label{pdot-eff} \d p & = \f{2 p (1 + \psi)}{\ga \sqrt\Delta \lp} \sin \b\mu c
\cos \b\mu c, \\
\label{cdot-eff} \d c & = -\f{3a}{2 \ga \Delta \lp^2} \sin^2 \b\mu c
+ \f{c(1 + \psi)}{\ga \sqrt\Delta \lp} \sin \b\mu c \cos \b\mu c
- \f{2 \pi G \ga}{a^5} (1 + 6 \psi) \pi_\vp^2.
\end{align}
Once again all nonlocal contributions due to a variation of $\b p$ vanish, as one should
expect.  Note that while classically $c$ is proportional to $\d p$, this is not the case
in the effective theory and the relation between them, as seen in Eq.\ \eqref{pdot-eff},
is considerably more complicated.  Finally, the relationship between $\pi_\vp$ and $\rho$
given in Eq.\ \eqref{dyn-rho} continues to hold.

It is interesting to solve for the Hubble rate by squaring Eq.\ \eqref{pdot-eff} and using
Eq.\ \eqref{sc-eff}, this gives the modified Friedmann equation (including perturbations),
\be \label{hubble} H^2 = \left( \f{\d p}{2p} \right)^2 = (1 + 2\psi) \f{8 \pi G}{3} \rho
\left( 1 - \f{\rho}{\rho_c} \right) - \f{2}{3a^2} \left( 1 - \f{2 \rho}{\rho_c} \right)
\na^2 \psi, \ee
where the critical energy density is $\rho_c = 3 / (8 \pi G \ga^2 \Delta \lp^2)$.  Clearly, the
classical result is obtained so long as the local matter energy density remains well below the
critical energy density.  It is also clear that when the matter energy density approaches the
Planck scale, the Hubble rate diminishes due to the repulsive nature of the quantum gravity
corrections, reaches zero close to the critical density (depending on the local strength of
the fluctuations in $\psi$) at which point there is a bounce where the Hubble rate begins to
grow again and as soon as the matter energy density drops below the Planck regime, contributions
due to quantum gravity are completely negligible.

Finally, the form of the equations of motion given above is a little unwieldy and for some
applications it will be useful to expand these equations and separate the background terms
from the perturbed quantities.   For the scalar field, this is easy,
\begin{align}
\label{bphidot-eff} \d{\b \vp} & = \f{\b\pi_\vp}{a^3 \sqrt{\oq}}, \\
\label{dphidot-eff} \d{(\de \vp)} & = \f{4 \psi \b\pi_\vp}{a^3 \sqrt{\oq}}
+ \f{\de \pi_\vp}{a^3 \sqrt{\oq}}; \\
\label{bpidot-eff} \d{\b \pi}_\vp & =  0, \\
\label{dpidot-eff} \d{(\de \pi_\vp)} & =  \sqrt{\oq}\, a \cos 2 \b\mu c \na^2 \de \vp;
\end{align}
but it becomes a little more involved for the geometric degrees of freedom.  For $p$,
since $p = a^2(1 - 2 \psi)$, we find that
\be \d\psi = \f{\d a}{a}(1 - 2 \psi) - \f{\d p}{2 a^2}. \ee
It is also necessary to expand the trigonometric functions of $\b\mu c$.  This gives,
for example,
\be \sin \b\mu c = \sin \f{\sqrt\Delta \lp \b c}{a} + \f{\sqrt\Delta \lp}{a} [\b c \psi
+ \de c] \cos \f{\sqrt\Delta \lp \b c}{a}. \ee
From these equations, it follows that
\begin{align}
\label{bpdot-eff} \d{\b p} & = 2 a \d a = \f{a^2}{\ga \sqrt\Delta \lp}
\sin \f{2 \sqrt\Delta \lp \b c}{a}, \\
\label{dpdot-eff} \d \psi & = \f{-\psi}{2 \ga \sqrt\Delta \lp}
\sin \f{2 \sqrt\Delta \lp \b c}{a}
- \f{\b c \psi + \de c}{\ga a} \cos \f{2 \sqrt\Delta \lp \b c}{a}; \\
\label{bcdot-eff} \d{\b c} & = \f{-3 a}{2 \ga \Delta \lp^2} \sin^2 \f{\sqrt\Delta \lp \b c}{a}
+ \f{\b c}{2 \ga \sqrt\Delta \lp} \sin \f{2 \sqrt\Delta \lp \b c}{a}
- \f{2 \pi G \ga}{a^5} \b\pi_\vp^2, \\
\label{dcdot-eff} \d{(\de c)} & = \f{\b c (\b c \psi + \de c)}{\ga a} \cos \f{2 \sqrt\Delta \lp \b c}{a}
- \f{\b c \psi + \de c}{\ga \sqrt\Delta \lp} \sin \f{2 \sqrt\Delta \lp \b c}{a}
- \f{12 \pi G \ga \psi}{a^5} \b\pi_\vp^2 -\f{4 \pi G \ga}{a^5} \b\pi_\vp \de\pi_\vp.
\end{align}
The trigonometric identities $2 \sin\theta \cos\theta = \sin 2\theta$ and $\cos^2 \theta
- \sin^2 \theta = \cos 2 \theta$ have been used in order to shorten the expressions above.

The dynamics, both of the background and of the perturbations around it, are given by the
equations above and one can check that, on-shell, $\d \mH^\eff = \d \mH_a^\eff = 0$, just as
one would expect.  Note that the $\cos 2 \b\mu c$ term that was added to the effective scalar
constraint in Eq.\ \eqref{scal-hol} is essential for $\d \mH^\eff = 0$ to hold.  This result
shows that the equations of motion and the constraints are consistent with each other.

\subsection{Conjectured Generalization to Other Perfect Fluids}
\label{ss.eff-fluid}

In this part, we provide a conjectured extension of our results to all other matter fields that
can be treated as perfect fluids.  It would be possible to use a Hamiltonian framework for
perfect fluids \cite{schutz}, but we will show that this is not necessary: the generalization
is straightforward enough that it can be done by hand.  There is a possibility that a
strong quantum backreaction from the matter field could spoil the validity of the
effective equations, but we will not consider this potential effect here.

It is worth pointing out that it would be significantly more difficult to generalize these
equations in order to include matter fields that allow anisotropic stress since at the very
beginning it was assumed that the perturbation in the lapse, $\phi$, was equal to the
perturbation in the scale factor $\psi$.  If this assumption is not made, then it is not
immediately clear how to construct a Hamiltonian formulation with the symplectic
structure given in Eq.\ \eqref{poisson}, without expanding it in terms of the background
and perturbations which would ruin the simple substitution used in Eq.\ \eqref{subF}.

Nonetheless, this generalization to all perfect fluids will be very useful, especially since
many matter fields behave like radiation at very high energies, which are precisely the
conditions to be expected when the curvature is near the Planck scale.

For generic perfect fluids, the effective scalar constraint becomes
\be \label{ham-gen} \rho - \f{3}{8 \pi G} \left[ \f{1}{\ga^2 \Delta \lp^2} \sin^2 \b\mu c
+ \f{2}{3 a^2} \na^2 \psi \right] = 0, \ee
which, as $\sin^2 \theta \le 1$, shows that the matter energy density is bounded above
by the critical density
\be \rho_c = \f{3}{8 \pi G \ga^2 \Delta \lp^2}, \ee
(modulo some small corrections from $\na^2 \psi$), which is consistent with Eq.\ \eqref{hubble}.

As before, the diffeomorphism constraint is unchanged and therefore
\be \label{diff-gen} \f{1}{\ga} \Big[ \b c \del_a \psi + \del_a (\de c) \Big]
- 4 \pi G a (\b\rho + \b P) \del_a (\de u) = 0.  \ee
Note however that since $c$ is not proportional to $\d p$ when holonomy corrections are
present, the diffeomorphism constraint cannot be simplified as much as it usually is when
there are no holonomy corrections.

The equations of motion for the matter degrees of freedom are easy to generalize.  Using
Eqs.\ \eqref{dyn-rho} and \eqref{pidot-eff}, along with the scalar constraint in order to
solve for the $\cos 2 \b\mu c$ term, we find that
\be \label{rhodot-gen} \d{\b \rho} + 3 \f{\d a}{a} (\b\rho + \b P) + \d{\de\rho} + 3 \f{\d a}{a}
(\de\rho + \de P) - 3 (\b\rho + \b P) \d\psi
+ \f{1}{a^2} (\b\rho + \b P) \na^2 (\de u) \left( 1 - \f{2 \rho}{\rho_c} \right) = 0, \ee
where the critical density $\rho_c$ appears once more.  The equation for $\d {\de u}$,
\be \label{udot-gen} \del_t \Big[ (\b\rho + \b P) \de u \Big] = -\de P - (\b\rho + \b P) \psi
- 3 \f{\d a}{a} (\b\rho + \b P) \de u, \ee
continues to hold.

Moving on to the geometrical degrees of freedom, since there are no matter terms in the equation
for $\d p$ it is trivially generalized,
\be \label{pdot-gen} \d p = \f{2 p (1 + \psi)}{\ga \sqrt\Delta \lp} \sin \b\mu c
\cos \b\mu c, \ee
while the generalization of Eq.\ \eqref{cdot-eff} is a little trickier.  A careful analysis of the
equation of motion for $\d c$ at the classical level shows that the $\pi_\vp^2$ appearing there is
related to the pressure of the massless scalar field rather than its energy density.  Therefore,
it follows that for an arbitrary perfect fluid,
\be \label{cdot-gen} \d c = -\f{3a}{2 \ga \Delta \lp^2} \sin^2 \b\mu c
+ \f{c(1 + \psi)}{\ga \sqrt\Delta \lp} \sin \b\mu c \cos \b\mu c - 4 \pi G \ga a P. \ee
Depending on the situation, it may be useful to expand these equations as was done in the previous
subsection in Eqs.\ \eqref{bphidot-eff}---\eqref{dcdot-eff}; this can be done by following the
same procedure that was used there.  Also, one can check that for these effective equations,
generalized in order to include all perfect fluids, the scalar and diffeomorphism constraints
are again preserved by the dynamics.

We can obtain the modified Friedmann equation by squaring Eq.\ \eqref{pdot-gen} and using Eq.\
\eqref{ham-gen},
\be \label{fried-gen} \f{\d a^2}{a^2} = \f{8 \pi G}{3} \rho \left( 1 - \f{\rho}{\rho_c} \right)
+ 2 \f{\d a^2}{a^2} \psi + 2 \f{\d a}{a} \d\psi
- \f{2}{3a^2} \na^2 \psi \left( 1 - \f{2 \rho}{\rho_c} \right), \ee
where it is understood that $\rho = \b\rho + \de\rho$ and the critical density $\rho_c$ appears again.
Similarly, the modified Raychaudhuri equation can be obtained by taking the time derivative of
Eq.\ \eqref{pdot-gen} and then using Eqs.\ \eqref{ham-gen}, \eqref{pdot-gen} and \eqref{cdot-gen} in
order to remove all of the trigonometric functions.  This gives
\be \label{raych-gen} \f{\ddot{a}}{a} = \f{\d a^2}{a^2} + 2 \f{\ddot{a}}{a} \psi
- 2 \f{\d a^2}{a^2} \psi + \f{\d a}{a} \d\psi + \ddot{\psi}
- \bigg[ 4 \pi G(\rho + P) - \f{1}{a^2} \na^2 \psi \bigg]
\bigg( 1 - \f{2 \rho}{\rho_c} + \f{1}{2 \pi G \rho_c a^2} \na^2 \psi \bigg). \ee
These equations can be compared with the usual general relativity results obtained in Sec.\ \ref{ss.dyn},
the classical results are obtained in the limit $\rho_c \to \infty$.

Any initial data satisfying Eqs.\ \eqref{ham-gen} and \eqref{diff-gen} can then be evolved by
using Eqs.\ \eqref{rhodot-gen}, \eqref{udot-gen}, \eqref{fried-gen} and \eqref{raych-gen}; these
last four equations are written in a form similar to the standard cosmological notation.  Although
the constraints are written in terms of different variables, one can use the standard classical
constraints given in Eqs.\ \eqref{ham-std} and \eqref{diff-std} in order to constrain the initial
data so long as the energy density $\rho$ is far from the critical density $\rho_c$ at the time
the initial conditions are set.  If the initial conditions are chosen at a time where $\rho$ is
within two orders of magnitude of $\rho_c$ or less, then the holonomy-corrected constraints
\eqref{ham-gen} and \eqref{diff-gen} must be used.

\section{Some Comments on Inverse Triad Corrections}
\label{s.inv}

In this section, we will make some comments regarding the type of inverse triad corrections that
might be expected in the effective equations for cosmological perturbations.  This section stands
apart from the remainder of the paper and can be skipped on a first reading.

There are two main inputs from loop quantum gravity that are used in order to construct the
Hamiltonian constraint operator in LQC.  First, the field strength of the Ashtekar connection
must be expressed in terms of holonomies and second, inverse triad operators must be introduced
carefully as the operator $\h p$ includes zero in its discrete spectrum.  In the previous
section, we showed how it is possible to treat holonomy corrections in an effective action; in
this section we shall comment on some of the properties that we expect inverse triad corrections
to have.  This has already been studied in LQC (see, e.g., the appendix of \cite{bck} where similar
considerations arise in a slightly different setting) but there are some important new aspects
which appear in the context of lattice LQC which provide further insights into this problem.

As is well known, there are many different ways to build inverse triad operators in LQC, all of
which have the correct semiclassical limit but behave differently at the Planck scale.  Because
of this ambiguity, it is not clear which inverse triad operator to choose.  Although there are
some heuristic arguments indicating that inverse triad corrections might be comparable to
holonomy corrections \cite{bct}, all of the inverse triad operators defined in LQC so far
vanish in noncompact space-times.  This indicates that our current understanding of inverse
triad operators is flawed as these operators should be local, rather than global, operators.
However, the inverse triad operators that naturally arise in the setting of lattice LQC have
different properties and may explain what the correct way to implement inverse triad operators
in a cosmological setting is.

In order to gain an idea of the form inverse triad corrections should take, we will consider
a relatively simple lattice LQC model \cite{boj-inh, we-inh}.
In this setting, we discretize the 3-torus into $N^3$ cells which are each taken to be homogeneous,
but as the gravitational and matter fields can vary from one cell to the next, inhomogeneities are
present at large scales.  The number of cells, $N^3$, gives the largest wave number perturbations
may have in this discretization,
\be \label{max-k} k = \f{N}{2}. \ee
Since the gravitational field is taken to be homogeneous in each cell, the induced symplectic
structure from the full theory changes and the Poisson brackets become
\be \{ c(\v n), p(\v m) \} = \f{8 \pi G \ga N^3}{3} \de_{\v n, \v m}, \ee
\be \{ \vp(\v n), \pi_\vp(\v m) \} = N^3 \de_{\v n, \v m}, \ee
where $\v n, \v m$ are vectors which label the cells in the three-dimensional lattice and
$\de_{\v n, \v m}$ is a Kronecker delta.  Note the explicit dependence of $N^3$ in the symplectic
structure, this will reappear in the inverse triad operators.

Following the full theory, inverse triad operators are constructed by starting with a Poisson
bracket (which is equal to $1/p$ when it is evaluated) between two functions on the phase space
that can easily be promoted to operators and then the inverse triad operator is given by the
commutator between the two operators, divided by $i \hbar$ \cite{tt}.  For lattice LQC, a
simple choice for the inverse triad operator corresponding to $1/p$ acting on a particular
cell is given by%
\footnote{This can be derived by noting that $p^{-1} = -(3i/4 \pi G \ga \sqrt{\Delta} \lp N^3)
e^{-i \b\mu c/2} \{ e^{i \b\mu c}, \sqrt{|p|} \} e^{-i \b\mu c/2}$ (where we have suppressed
the label $\v n$ in order to simplify the notation) and then promoting the right side to be
an operator by replacing $\{ \cdot, \cdot \} \to [ \cdot, \cdot ] / i\hbar$.  The action of
this operator is most easily seen by changing variables to $\nu \propto p^{3/2}$ so that
$e^{i \b \mu c}$ acts as a simple shift operator on states $|\nu\ket$.  Then, changing
variables back to $p$ one obtains the result given in Eq. \eqref{invop}.  See
\cite{as-rev, aps} for further information on inverse triad operators in the $\b\mu$
scheme in homogeneous and isotropic LQC.}
%
\begin{align} \label{invop} \wh {\f{1}{p(\v n)}} | p(\v n) \ket =
\f{3}{4 \pi \ga \sqrt{\Delta} \lp^3 N^3}
\bigg[ & \left| |p(\v n)|^{3/2} + 2 \pi \ga \sqrt\Delta \lp^3 N^3 \right|^{1/3} \nn \\ &
- \left| |p(\v n)|^{3/2} - 2 \pi \ga \sqrt\Delta \lp^3 N^3 \right|^{1/3} \bigg]
| p (\v n) \ket. \end{align}
Clearly this operator is always well-defined, even on the state $| p(\v n) = 0 \ket$, which it
annihilates.  Also, the eigenvalue approximates $1/p$ for large $p$.

In order to incorporate inverse triad corrections for the operator chosen above, one would simply
replace all occurances of $1/p$ in the scalar constraint with the right hand side of Eq.\ \eqref{invop}.
Since we know that the singularity is avoided due to the holonomy corrections, it follows that
$p \neq 0$ and then inverse triad corrections can be written as
\be \f{1}{p} \to \f{1}{p} \times F(p, N), \ee
where the corrections are encoded in the function $F(p, N)$, 
\be F(p, N) = \f{3}{2 f(p, N)} \left[ |1 + f(p, N)|^{1/3} - |1 - f(p, N)|^{1/3} \right], \ee
and $f(p, N)$, in turn, is given by
\be f(p, N) = \f{2 \pi \ga \sqrt\Delta \lp^3 N^3}{p^{3/2}}. \ee

Since the inverse triad correction depends on the number of cells $N^3$, this type of correction
is sometimes viewed as being unphysical.  However, since the physical wavelength of the Fourier
mode in a compact, (approximately) homogeneous, space of volume $p^{3/2}$ divided into $N^3$ cells
is $\la_{\rm phy} = 2\sqrt{p}/N$ [see Eq.\ \eqref{max-k}], it follows that $F$ and $f$ should not
be viewed as functions of $p$ and $N$ but rather of the physical wavelength of the Fourier mode
studied, $\la_{\rm phy}$.  Then,
\be F(\la_{\rm phy}) = \f{3}{2 f(\la_{\rm phy})} \left[ |1 + f(\la_{\rm phy})|^{1/3}
- |1 - f(\la_{\rm phy})|^{1/3} \right], \ee
\be f(\la_{\rm phy}) = \f{16 \pi \ga \sqrt\Delta \lp^3}{\la_{\rm phy}^3}. \ee
\emph{The inverse triad correction to a given Fourier mode depends on the ratio of the wavelength
of that mode to the Planck length}.  For modes with a wavelength considerably larger than $\lp$,
inverse triad corrections are completely negligible, even if the space-time curvature is large.

The dependence of the inverse triad correction on the wavelength of each mode is completely natural.
In homogeneous models in LQC, inverse volume effects are only relevant in compact space-times and then
the strength of the correction depends on the ratio of the volume of the space-time to the Planck
volume.  This indicates that for inverse triad effects to be present, there must be length scales in
the space-time which can be compared to the Planck length.  In homogeneous space-times there is only
the size of the entire space-time which provides a length scale%
\footnote{This is why inverse volume corrections in homogeneous, noncompact space-times vanish in LQC.},
but when perturbations are present there are additional length scales provided by the wavelengths of the
Fourier modes of the perturbations.  Therefore, this property of the inverse triad corrections should
not be surprising.

As a final note, we point out that much of the previous work studying inverse triad corrections in
the cosmology perturbation equations for scalar, vector and tensor modes do not allow for the
inverse triad corrections to depend on the wavelength of the mode, instead the corrections typically
only depend on the volume of the background homogeneous space-time.  We suggest that these results
should be generalized.

\section{Discussion}
\label{s.dis}

We have shown how it is possible to incorporate holonomy-type corrections into the equations of
motion for perturbations around a flat FLRW background.  This was first done in the case of a
massless scalar field and then we presented a conjectured generalization for all perfect fluid
matter fields.

Since holonomy corrections are nonperturbative, we decided to use elementary variables that are
nonperturbative.  This is to say that they are not separated into background and perturbation
terms, rather it is after the equations of motion are obtained that this split is performed.
It was then necessary to work in the longitudinal gauge so that we could restrict our attention
to holonomies that are almost periodic in the connection.  These two choices greatly simplified
the problem  and the correct way to implement holonomy corrections in the effective Hamiltonian
constraint for this particular gauge became more obvious.

It would be nice to extend this analysis in order to include tensor and vector modes, but this
may not be as trivial as one would hope.  The problem is that a lot of the simplifications which
made this problem tractable at a nonperturbative level ---such as a diagonal metric which allowed
us to work with a relatively small phase space and restrict our attention to holonomies that are
almost periodic functions of the connection--- are not possible once tensor and vector perturbations
are allowed.  Therefore it seems that a combination of the techniques used here and perturbative
methods used in \cite{bh-v, bh-t, mcbg, bhks1, bhks2, wl} will be necessary in order to be able
to treat scalar, vector and tensor perturbations all together.  Despite these remaining open
issues, a lot can be learnt from this treatment of scalar perturbations.

First, the diffeomorphism and Gauss constraints are unchanged and therefore the symmetries they
encode hold exactly at all scales, including the Planck scale.  On the other hand, the scalar
constraint and the dynamics of the gravitational field are modified by the holonomy corrections.
As we saw in Sec.\ \ref{s.eff}, the holonomy corrections are introduced in precisely the same way
as in the homogeneous models and the resulting constraints are shown to be preserved by the
dynamics.  As usual in LQC, we found that the holonomy corrections are parametrized by the critical
energy density $\rho_c$; these corrections vanish in limit of $\rho_c \to \infty$ and then the
standard cosmological perturbation equations are recovered.

Although it is not yet clear how to include inverse triad operators ---which are essential if
one wants to construct the quantum theory--- it is possible to learn some lessons from a
discretization of LQC.  The main point is that the inverse triad corrections should vary from
one Fourier mode of the perturbation to the next.  To be more precise, the inverse triad
corrections, for a given Fourier mode of the perturbation, should explicitly depend upon the
ratio of the wavelength of the mode to the Planck length.  Thus, the shorter the wavelength,
the more important the inverse triad correction will be.  This is natural as the only physical
length scale provided by perturbations is the wavelength of the modes.  Because of this,
inverse triad corrections should be expected to be completely negligible so long as the
wavelengths of the perturbative modes of interest stay an order of magnitude or two away from
the Planck scale.  However, if the wavelengths of interest approach the Planck scale, then
inverse triad corrections cannot be ignored.  For example, these corrections could be important
in inflationary scenarios, but not in ekpyrotic or matter bounce models.

This shows that holonomy corrections and inverse triad corrections have a completely different
nature: holonomy corrections become important when the space-time curvature approaches the
Planck scale while inverse triad operators become important when physical length scales are
comparable to the Planck length.  Thus, depending upon the situation, one of the corrections
may be important while the other is negligible.  In particular, we expect the equations derived
in this paper hold at all curvature scales so long as the wavelengths of the scalar modes of
interest remain large compared to $\lp$.

\acknowledgments

This work was supported by Le Fonds qu\'eb\'ecois de la recherche sur la nature
et les technologies.

\end{document}